\documentclass[useAMS,usenatbib]{mn2e}

\usepackage[]{color,graphicx}
\usepackage[bookmarks=true,bookmarksnumbered=true,
 pdftitle={bulk},%
 pdfauthor={Yudai Suwa},%
 pdfkeywords={TeX; dvipdfmx; hyperref; color;},%
 colorlinks=true, %
 citebordercolor={0 0 1}]
  {hyperref}

\usepackage{amsmath}

\def\lesssim{\mathrel{\hbox{\rlap{\hbox{\lower5pt\hbox{$\sim$}}}\hbox{$<$}}}}
\def\gtrsim{\mathrel{\hbox{\rlap{\hbox{\lower5pt\hbox{$\sim$}}}\hbox{$>$}}}}

\title[How Much Can $^{56}$Ni Be Synthesized by Magnetar Model?]
{How Much Can $^{56}$Ni Be Synthesized by Magnetar Model for Long
  Gamma-ray Bursts and Hypernovae?}

\author[Y. Suwa and N. Tominaga]{
Yudai Suwa$^{1,2}$\thanks{E-mail: suwa@yukawa.kyoto-u.ac.jp}
and
Nozomu Tominaga$^{3,4}$
\\
$^{1}$Yukawa Institute for Theoretical Physics, Kyoto University,
Oiwake-cho, Kitashirakawa, Sakyo-ku, Kyoto, 606-8502, Japan\\
$^{2}$Max-Planck-Institut f\"ur Astrophysik,
Karl-Schwarzschild-Str. 1, D-85748 Garching, Germany\\
$^{3}$Department of Physics, Faculty of Science and
Engineering, Konan University, 8-9-1 Okamoto, Kobe, Hyogo 658-8501,
Japan\\
$^{4}$Kavli Institute for the Physics and Mathematics of
the Universe (WPI), The University of Tokyo, Kashiwa, Chiba
277-8583, Japan
}

\begin{document}

\date{Accepted. Received.}

\pagerange{\pageref{firstpage}--\pageref{lastpage}} \pubyear{2015}

\maketitle

\label{firstpage}

\begin{abstract}
A rapidly rotating neutron star with strong magnetic fields, called
magnetar, is a possible candidate for the central engine of long
gamma-ray bursts and hypernovae (HNe). We solve the evolution of a
shock wave driven by the wind from magnetar and evaluate the
temperature evolution, by which we estimate the amount of $^{56}$Ni
that produces a bright emission of HNe. We obtain a constraint on the
magnetar parameters, namely the poloidal magnetic field strength
($B_p$) and initial angular velocity ($\Omega_i$), for synthesizing
enough $^{56}$Ni mass to explain HNe ($M_{^{56}\mathrm{Ni}}\gtrsim
0.2M_\odot$),
i.e. $(B_p/10^{16}~\mathrm{G})^{1/2}(\Omega_i/10^4~\mathrm{rad~s}^{-1})\gtrsim
0.7$.
\end{abstract}


\section{Introduction}
\label{sec:intro}

The central engine of gamma-ray bursts (GRBs) is still unknown
nevertheless a wealth of observational data. The most popular scenario
for a subclass with long duration (long GRB) is the collapsar scenario
\citep{woos93}, which contains a black hole and a hyper accretion
flow, and one of the alternatives is a rapidly rotating neutron star
(NS) with strong magnetic fields (``magnetar'') scenario
\citep{usov92}. Their energy budgets are determined by the
gravitational binding energy of the accretion flow for the former
scenario and the rotational energy of a NS for the latter scenario.

On the other hand, the association between long GRBs and energetic
supernovae, called hypernovae (HNe), is observationally established
since GRB 980425/SN 1998bw and GRB~030329/SN~2003dh \citep[see][and
  references therein]{woos06,hjor12}. The explosion must involve at
least two components; a relativistic jet, which generates a gamma-ray
burst, and a more spherical-like non-relativistic ejecta, which is
observed as a HN. One of observational characteristics of HNe is high
peak luminosity; HNe are typically brighter by $\sim1-2$~mag than
canonical supernovae. The brightness of HNe stems from an ejection of
a much larger amount of $^{56}$Ni (0.2 -- 0.5 $M_\odot$;
\citealt{nomo06}) than canonical supernovae ($\lesssim 0.1 M_\odot$,
e.g., \citealt{blin00} for SN~1987A).

Mechanisms that generate such a huge amount of $^{56}$Ni by a HN have
been investigated
\citep[e.g.][]{macf99,naka01b,naka01c,maed02,naga06,tomi07,maed09}.
They demonstrated that the large amount of $^{56}$Ni can be
synthesized by explosive nucleosynthesis due to the high explosion
energy of a HN and/or be ejected from the accretion disk via disk
wind. On the other hand, no study on the $^{56}$Ni mass for the
magnetar scenario has been done so far. The dynamics of outflow from
magnetar is investigated in detail and it is suggested that the energy
release from the magnetar could explain the high explosion energy of
HNe \citep[e.g.][]{thom04,komi07,dess08,bucc09,metz11}. 
Not only the explosion dynamics, but also self-consistent evolutions
of magnetized iron cores have been investigated for more than four
decades
\citep[e.g.,][]{lebl70,meie76,symb84,burr07b,wint12,sawa13,moes14,nish15},
in which magneto-hydrodynamic (MHD) equations were solved.  In these
simulations, rapidly rotating ($P\sim O(1)$ s) and strongly magnetized
($B\sim 10^{9-12}$ G) cores are employed as initial conditions. The
final outcomes after the contraction of cores to NSs are very rapidly
rotating ($P\sim O(1)$ ms) and very strongly magnetized ($B\sim
10^{14-16}$ G) NSs, which can generate magnetic-driven outflows.
These studies, however, basically focused on the shock dynamics
affected by strong magnetic fields and/or yield of r-process elements,
but have scarcely paid attention to $^{56}$Ni amount so far.
Additionally, these simulations have not been able to produce strong
enough explosion explaining HNe, but trying to explain canonical
supernovae (the explosion energy $\sim 10^{51}$ erg; for HNe $\sim
10^{52}$ erg is necessary).
Therefore, there is a need to study the amount of $^{56}$Ni generated
by the magnetar central engine in order to check the consistency of
this scenario.

In this paper, we evaluate the amount of $^{56}$Ni by the rapidly
spinning magnetar. To do this, we adopt a thin shell approximation and
derive an evolution equation of a shock wave driven by the magnetar
dipole radiation. The solution of this equation gives temperature
evolution of post-shock layer. Using the critical temperature
($5\times 10^9$ K) for nuclear statistical equilibrium at which
$^{56}$Ni is synthesized, we give a constraint on the magnetar spin
rate and dipole magnetic field strength for explaining the
observational amount of $^{56}$Ni in HNe.  In Section
\ref{sec:method}, we give expressions for the dipole radiation from a
rotating magnetized NS for the central engine model and the derivation
of the evolution equation of a shock wave.  Based on the solution, we
evaluate the temperature evolution and $^{56}$Ni mass
($M_{^{56}\mathrm{Ni}}$) as a function of magnetar parameters in
Section \ref{sec:results}. We summarize our results and discuss their
implications in Section \ref{sec:summary}.

\section{Computational Method}
\label{sec:method}

According to \citet{shap83}, the luminosity of dipole radiation is
given as
\begin{equation}
L_w=\frac{B_p^2R^6\Omega^4\sin^2\alpha}{6c^3},
\end{equation}
where $B_p$ is the dipole magnetic filed strength, $R$ is the NS
radius, $\Omega$ is the angular velocity, $\alpha$ is the angle
between magnetic and angular moments, and $c$ is the speed of
light. Hereafter we assume $\sin\alpha=1$ for simplicity. Then, the
luminosity is expressed as
\begin{eqnarray}
L_w&=&6.18\times 10^{51}\mathrm{erg~s^{-1}}\nonumber\\
&&\times
\left(\frac{B_p}{10^{16}~\mathrm{G}}\right)^2
\left(\frac{R}{10~\mathrm{km}}\right)^6
\left(\frac{\Omega}{10^4~\mathrm{rad~s^{-1}}}\right)^4.
\label{eq:Lw}
\end{eqnarray}
The time evolution of the angular velocity is given as
\begin{eqnarray}
\Omega(t)&=&\Omega_i
\left(1+\frac{t}{T_d}\right)^{-1/2},
\end{eqnarray}
where $\Omega_i$ is the initial angular velocity and $T_d$ is spin
down timescale given by
\begin{eqnarray}
T_d&=&\frac{3Ic^3}{B_p^2R^6\Omega_i^2}\nonumber\\
&=&8.08~\mathrm{s}
\left(\frac{B_p}{10^{16}~\mathrm{G}}\right)^{-2}
\left(\frac{R}{10~\mathrm{km}}\right)^{-6}\nonumber\\
&&\times\left(\frac{\Omega_i}{10^4~\mathrm{rad~s^{-1}}}\right)^{-2}
\left(\frac{I}{10^{45}~\mathrm{g~cm^2}}\right),
\label{eq:Td}
\end{eqnarray}
where $I$ is the moment of inertia of a NS. Therefore, $L_w(t)\propto
(1+t/T_d)^{-2}$. The available energy is the rotation energy of a NS,
\begin{equation}
E_\mathrm{NS}=\frac{1}{2}I\Omega_i^2
=5\times 10^{52}~\mathrm{erg}
\left(\frac{I}{10^{45}~\mathrm{g~cm^2}}\right)
\left(\frac{\Omega_i}{10^4~\mathrm{rad~s}^{-1}}\right)^2,
\label{eq:ENS}
\end{equation}
which corresponds to the total radiation energy $E_w=\int_0^\infty
L_w(t)dt=L_w(0)T_d$.

Next, we calculate the time evolution of the shock. For simplicity, we
employ thin shell approximation for the ejecta
\citep[e.g.,][]{laum69,koo90,whit02}. In this picture, we consider an
isotropic wind, which forms a hot bubble. This bubble sweeps up the
surrounding matter into a thin dense shell. This approximation is
applicable when the thickness between forward and reverse shocks is
small compared to their radii.  The comparisons of our solutions with
hydrodynamic simulations are shown in Appendix.

The equation of motion of the shell is given as
\begin{equation}
\frac{d}{dt}\left(M_s \dot R_s\right)=4\pi R_s^2 p-F_g,
\label{eq:EOM}
\end{equation}
where $R_s$ is the shock radius, $M_s$ is mass of the shell, and $p$
is the pressure below the shell, which drives the shell. $F_g$ is the
gravitational force, which consists of contributions from a point
source ($GM_cM_s/R_s^2$; $G$ is the gravitational constant and $M_c$
is the mass below the shell) and the self gravity ($GM_s^2/2R_s^2$).
$\dot R_s$ denotes the derivative of $R_s$ with respect to time. The
left hand side (LHS) represents the increase rate of the outward
momentum, while the first term of the right hand side (RHS) is the
driving force of the shell propagation due to the pressure $p$.  We
neglect the ram pressure in this model because the ram pressure of the
falling matter does not affect on the evolution of the shock after the
onset \cite[e.g.,][]{tomi07}. However, since the ram pressure is
highest at the onset of the propagation and influences on the onset,
we take into account the effect with a condition that the shock
propagation time should be shorter than the free-fall
time.\footnote{In order to onset the shock propagation, the ram
  pressure of the falling matter $\rho v_\mathrm{ff}^2$ is overwhelmed
  by the thermal pressure $p$. According to Eq. (\ref{eq:EOM}), the
  thermal pressure is $p\sim\dot{R_s} \dot{M_s}/4 \pi R_s^2$ and the
  ram pressure is $\rho v_\mathrm{ff}^2\sim v_\mathrm{ff} \dot{M} / 4
  \pi R_s^2$, where $\dot M\sim 4\pi R_s^2 \rho v_\mathrm{ff}$.  Thus,
  the condition is $\dot{R}>v_\mathrm{ff}$.}  The ambiguity originated
from this approximation is checked by comparing evolutions of shock
and temperature with hydrodynamic simulations (see Appendix).

The energy conservation of the bubble is given as
\begin{equation}
\frac{d}{dt}\left(\frac{4\pi}{3}R_s^3\frac{p}{\gamma-1}\right)=L_w-p\frac{d}{dt}\left(\frac{4\pi}{3}R_s^3\right),
\label{eq:energy}
\end{equation}
where $\gamma$ is the adiabatic index and $L_w$ is the wind driven by
the magnetar, which is assumed to be the dipole radiation given by
Eq. (\ref{eq:Lw}). The term on the LHS is the increase rate of the
internal energy of the bubble, while terms on the RHS are the energy
injection rate by the wind and the power done by the bubble pushing on
the shell. Note that it is assumed that the other mechanisms, such as
neutrino heating, give no energy to the shock.

Nuclear statistical equilibrium holds and $^{56}$Ni is synthesized in
a mass shell with the maximum temperature of $>5\times 10^9$ K. Thus,
the temperature evolution is crucial for the amount of $^{56}$Ni. In
the following, we consider the postshock temperature, which is
evaluated with the following equation of state,
\begin{equation}
p=p_i+p_e+p_r,
\end{equation}
where $p_i=n_ik_BT$, $p_e=(7/12)a_\mathrm{rad}T^4[T_9^2/(T_9^2+5.3)]$,
and $p_r=a_\mathrm{rad}T^4/3$ are contributions from ions,
non-degenerate electron and positron pairs \citep{frei99,tomi09}, and
radiation, respectively. Here, $n_i=\rho/m_p$ is the ion number
density with $m_p$ being the proton mass and $\rho$ being the density
in the shell,\footnote{Note that $\rho$ should be different from
  $\rho_0$ because matter is compressed by the shock wave. Due to our
  simple thin shell approximation we need an additional assumption to
  evaluate $\rho$. We hereby simply assume that $\rho=\rho_0$, which
  would lead to higher temperatures. Although the pressure inside the
  shell might also be different from the one behind the shell, we
  neglect the difference for simplicity.} $T$ is the temperature in
the shell, $T_9=(T/10^9~\mathrm{K})$, $k_B$ is Boltzmann's constant,
and $a_\mathrm{rad}=7.56\times 10^{-15}$ erg cm$^{-3}$ K$^{-4}$ is the
radiation constant. Combined with Eq. (\ref{eq:EOM}), we obtain $T$ in
the shell and its evolution being consistent with the shock dynamics.

By substituting Eq. (\ref{eq:EOM}) into (\ref{eq:energy}) and eliminating
$p$, we get
\begin{eqnarray}
&&(3\gamma-4)GM_s(2M_c+M_s)\dot R_s+24\pi\gamma\rho_0 R_s^4\dot R_s^3\nonumber\\
&&+8\pi R_s^5\dot R_s(\rho_0'\dot R_s^2+3\rho_0\ddot R_s)\nonumber\\
&&-2R_s^2\left[3(\gamma-1)L_w-(3\gamma-2)M_s\dot R_s\ddot R_s\right]\nonumber\\
&&+2R_s^3\left[4\pi G(M_c+M_s)\rho_0\dot R_s+M_s\dddot R_s\right]=0,
\label{eq:evolution}
\end{eqnarray}
where $\rho_0(r)$ is the density of the progenitor star
(i.e. pre-shocked material) and $\rho_0'=d\rho_0/dr$. In this
calculation, we used $\dot M_s=dM_s/dt=(dR_s/dt)(dM_s/dR_s)=4\pi
R_s^2\rho_0(R_s) \dot R_s$.  Note that all mass expelled by the shell
is assumed to be accumulated in the shell.
For the density structure, $\rho_0$, we employ s40.0 model of
\citet{woos02}, which is a Wolf-Rayet star with a mass of 8.7$M_\odot$
and a radius of 0.33$R_\odot$. In addition, we use $\gamma=4/3$.
Eq. (\ref{eq:evolution}) can be written to as a set of first order
differential equations,
\begin{eqnarray}
R_0(t)&=&R_s(t),\\
\dot R_0(t)&=&R_1(t),\\
\dot R_1(t)&=&R_2(t),\\
\dot R_2(t)&=&f(R_0, R_1, R_2),
\end{eqnarray}
where
\begin{eqnarray}
&&f(R_0, R_1, R_2)=\nonumber\\
&&~~ -\frac{GR_1}{2M_sR_0^3}[(3\gamma-4)(2M_c+M_s)M_s+8\pi R_0^3\rho_0(M_c+M_s)]\nonumber\\
&&~~ -\frac{12\pi\gamma}{M_s}\rho_0R_0R_1^3-\frac{4\pi}{M_s}R_0^2R_1(\rho_0'R_1^2+3\rho_0R_2)\nonumber\\
&&~~ +\frac{1}{M_sR_0}[3(\gamma-1)L_w-(3\gamma-2)M_sR_1R_2].
\end{eqnarray}
This system of differential equations is integrated using the fourth
order Runge-Kutta time stepping method. These equations allow us to
investigate the shock propagation in the realistic stellar model,
which depends on the density structure and the evolution of the energy
injection.

\section{Results}
\label{sec:results}

\begin{figure}
\centering
\includegraphics[width=0.45\textwidth]{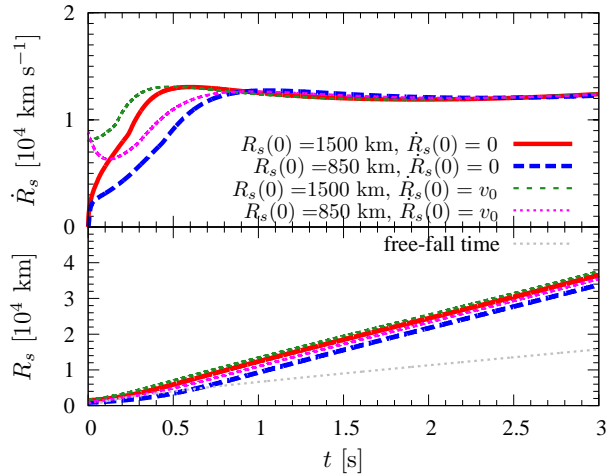}
\caption{Time evolutions of shock velocity (top panel) and shock
  radius (bottom panel). Four different lines represent different
  initial conditions for the shock radius ($R_s(0)=850$ km or 1500 km)
  and shock velocity ($\dot R_s(0)=0$ or
  $v_0=\sqrt{GM_c/2R_s(0)}$). The grey dotted line in the bottom panel
  represents the free-fall time at each radius.}
\label{fig:shock}
\end{figure}

Figure \ref{fig:shock} presents the time evolutions of shock radius
and shock velocity for a constant luminosity of $L_w=10^{52}$ erg
s$^{-1}$. Three boundary conditions are needed to solve
Eq. (\ref{eq:evolution}) because it is a third order differential
equation. We set $R_s$, $\dot R_s$, and $\ddot R_s$ at $t=0$. Figure
\ref{fig:shock} shows models with different initial conditions; models
with different injection points $R_s(t=0)=1500$ km ($M_c=1.5M_\odot$;
red thick-solid and green thin-dashed lines), and $R_s(0)=850$ km
($M_c=M_\odot$; blue thick-dashed and magenta thin-dotted lines), and
models with different initial velocity $\dot R_s(0)=0$ (two thick
lines) and $\dot R_s(0)=v_0\equiv\sqrt{GM_c/2R_s(0)}$ (two thin lines)
that is velocity necessary to overwhelm ram pressure
\citep[see][]{maed09}.  We find that the dependence on the initial
$\ddot R_s$, which is $0$ for all models shown in this figure, is very
minor so that we do not show its dependence here. In these
calculations, $M_s(t=0)=0$, i.e. the mass below $R_s(0)$ is assumed to
be a compact object and does not contribute to the mass of the shell.
The almost constant velocity is a consequence of the density
structure, $\rho_0(r)\propto r^{-\beta}$, with $\beta\approx 2$.  The
grey dotted line in the bottom panel represents the free-fall time
scale, $t_\mathrm{ff}=\sqrt{R_s^3/G(M_c+M_s)}$, for the corresponding
radius.

\begin{figure}
\centering
\includegraphics[width=0.45\textwidth]{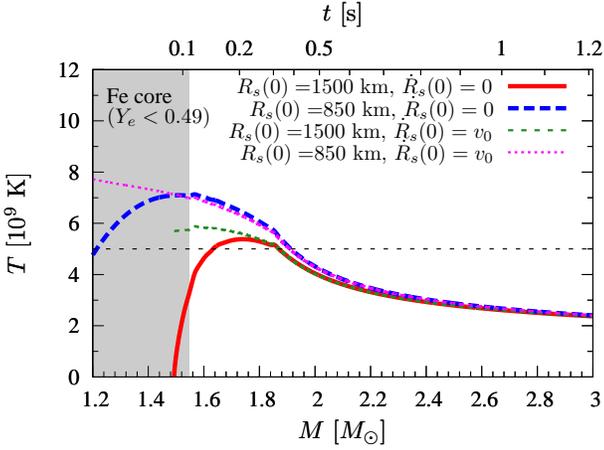}
\caption{The postshock temperature as a function of mass
  coordinate. The model parameters are the same as in Figure
  \ref{fig:shock}. The horizontal dotted line represents $5\times
  10^9$ K, above which $^{56}$Ni is synthesized. The gray shaded
  region, $M(r)<1.55M_\odot$, is the iron core, where $^{56}$Ni cannot
  be synthesized due to the low electron fraction of $Y_e<0.49$. The
  corresponding time of the model with $R_s(0)=850$ km and $\dot
  R_s(0)=v_0$ (magenta thin-dotted line) is given on the upper axis.}
\label{fig:temp}
\end{figure}

Figure \ref{fig:temp} gives the temperature in the expanding shell as
a function of mass coordinate for the same model as in Figure
\ref{fig:shock}. The electron fraction in the iron core ($M\lesssim
1.55 M_\odot$) is less than 0.49 so that no $^{56}$Ni production is
expected. The maximum temperature of each mass element is determined
by the energy injected until the shock front reaches the mass
element. Thus, in order to achieve $T>5\times 10^9$ K just above the
iron core, an initially fast shock wave or a shock injected deep
inside is necessary. This is because smaller initial velocity leads to
a smaller initial kinetic energy, and larger injection radius leads to
shorter and smaller energy injection before the shock reaches a
certain radius.  We employ $R_s(0)=850$ km and $\dot R_s(0)=v_0$ to
evaluate the {\it maximum} amount of $^{56}$Ni in the following
calculation. Although the model with $R_s(0)=850$ km and $\dot
R_s(0)=0$ represents similar temperature, its expansion time of the
shell is comparable to the free-fall time even for $L_w=10^{52}$ erg
s$^{-1}$ (see Fig. \ref{fig:shock}), so that the explosion might fail.

\begin{figure}
\centering
\includegraphics[width=0.45\textwidth]{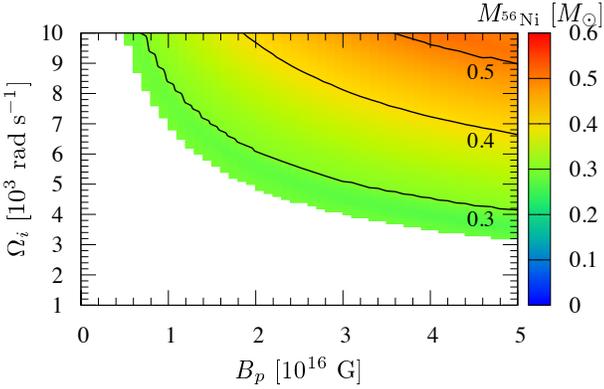}
\caption{The amount of $^{56}$Ni in units of $M_\odot$ for magnetar
  model as a function of the strength of the dipole magnetic field,
  $B_p$ and the initial angular velocity, $\Omega_i$. The region with
  $M<1.55 M_\odot$ is not included because $Y_e<0.49$ and no $^{56}$Ni
  production is expected there.  Black solid lines represent
  $M_{^{56}\mathrm{Ni}}$ from 0.3 to 0.5 $M_\odot$.}
\label{fig:e-t-m}
\end{figure}

Next, we consider the shock driven by the magnetar's dipole radiation.
Figure \ref{fig:e-t-m} shows the $^{56}$Ni mass produced in the
expanding shell as a function of $B_p$ and $\Omega_i$. In this figure,
we employ $R_\mathrm{NS}=10$ km and $I=10^{45}$ g cm$^2$. Here, we
assume that the matter that experienced $T>5\times10^9$~K is
completely converted to $^{56}$Ni, i.e., $X(^{56}{\rm Ni})=1$, except
for $M(r)<1.55 M_\odot$ where $Y_e<0.49$. From this figure, we can
easily see a rapid increase from 0 to $\sim$0.2$M_\odot$ of
$M_{^{56}\mathrm{Ni}}$.  In this progenitor, the silicon core has a
mass of $\sim 1.84M_\odot$, and the density slope $\beta$ is different
in the surrounding oxygen layer. This change in $\beta$ causes the
change of velocity evolution shown in Figure \ref{fig:shock}: for
instance, the blue thick-dashed line represents a rapid acceleration
at $t\lesssim 0.5$ s and a slow acceleration or an almost constant
velocity afterwards.

Since the observed brightness of HNe requires $\sim 0.2$ --
$0.5M_\odot$ of $^{56}$Ni \citep{nomo06}, a reasonable central engine
model must achieve this quantity.  We find that for
$M_{^{56}\mathrm{Ni}}\gtrsim 0.2M_\odot$, the following relation
should be satisfied;
\begin{equation}
\left(\frac{B_p}{10^{16~\mathrm{G}}}\right)^{1/2}
\left(\frac{\Omega_i}{10^4~\mathrm{rad~s^{-1}}}\right)
\gtrsim 0.68.
\label{eq:constraint}
\end{equation}
This condition can be derived by $E_\mathrm{NS}/T_d\gtrsim5.3\times
10^{50}$ erg s$^{-1}$ (see Eqs. \ref{eq:Td} and \ref{eq:ENS}).  Note
that Eq. (\ref{eq:constraint}) is a conservative constraint because in
this calculation we made several approximations, which always result
in larger $M_{^{56}\mathrm{Ni}}$. Thus, for a more realistic case,
$M_{^{56}\mathrm{Ni}}$ becomes smaller than this estimate. To make a
reasonable amount of $^{56}$Ni to explain the observation, a more
energetic central engine is needed.

In order to investigate the progenitor dependence, we perform the same
calculation with different progenitor models and find that the RHS of
Eq. (\ref{eq:constraint}) is $\sim$0.64 -- 0.90; 0.68 for 20
$M_\odot$, 0.90 for 40 $M_\odot$, 0.64 for 80 $M_\odot$ models of
\citet{woos07}, and 0.71 for 20 $M_\odot$ model of \citet{umed05}.
Therefore, this criterion does not strongly depend on the detail of
the progenitor structure. These calculations are performed with
$M_c=M_\odot$ and $\dot R_s(0)=v_0$.

\section{Summary and discussion}
\label{sec:summary}

In this study, we employed the thin shell approximation for shock
structure and calculated evolution of a shock wave driven by wind from
a rapidly rotating neutron star with strong magnetic fields
(``magnetar''). By evaluating temperature evolution that is consistent
with the shock evolution, we obtained a constraint on the magnetar
parameters, namely magnetic field strength and rotation velocity (see
Eq. \ref{eq:constraint}), for synthesizing enough amount of $^{56}$Ni
to explain brightness of HNe.

In this calculation, we employed several assumptions.
\begin{itemize}
\item The dipole radiation is dissipated between the NS and the shock
  and thermal pressure drives the shock evolution. This assumption
  leads to larger amount of $^{56}$Ni than more realistic situations
  because if the conversion from Poynting flux to thermal energy is
  insufficient, the internal energy is smaller and the temperature in
  the shell is lower than the current evaluation. Therefore, the mass
  that experienced $T>5\times 10^9$ becomes smaller.
\item The shock and energy deposition from the magnetar are spherical,
  which leads to larger $^{56}$Ni mass.  This is because fallback of
  matter onto a NS takes place and reduces $M_{^{56}\mathrm{Ni}}$, if
  the explosion energy is concentrated in a small region
  \citep{bucc09,maed09,yosh14}.
\item All energy radiated by the NS is used for HN component, which is
  overestimated because a part of the energy should be used to make
  the relativistic jet component of a GRB.
\item The density inside the shell is assumed to be the same as the
  progenitor model. This assumption results in the higher temperature
  and the larger $M_{^{56}\mathrm{Ni}}$ than realistic hydrodynamical
  calculations because the shock enhances not only the pressure but
  also the density in the shell.
\item Matter which experiences $T>5\times 10^9$K consists only of
  $^{56}$Ni, i.e. $X(^{56}\mathrm{Ni})=1$. This overestimates
  $M_{^{56}\mathrm{Ni}}$ because $X(^{56}\mathrm{Ni})<1$ even in the
  layer which experiences $T>5\times 10^9$ K according to
  hydrodynamical and nucleosynthesis simulations \citep{tomi07}.
\item The mass cut corresponds to the iron core mass, $1.55M_\odot$.
  If the NS mass is larger than the iron core mass, the $^{56}$Ni mass
  becomes even smaller.
\item The ram pressure is neglected in the evolutionary equation of
  the shell. According to the estimate of the shock propagation time
  and the free-fall time, in the low luminosity case the shell could
  not propagate outward for more realistic calculations.
\end{itemize}
Combining these facts, our estimation of the $^{56}$Ni mass is
probably highly overestimated so that our constraint on the magnetar
parameters (Eq. \ref{eq:constraint}) is rather conservative.
Interestingly, it is still a stringent constraint; a very high
magnetic field strength and a very rapid rotation are required to
explain the brightness of HNe.

Next, we discuss about more detailed MHD simulations for mechanisms
driving ejecta by transferring rotational energy of magnetars using
magnetic fields, although the mechanism is different from dipole
radiation assumed in this study.
\citet{bucc09} performed MHD simulations around new-born magnetars
from 1 s after supernova shock emergence and found that the energy
extracted from magnetars through magnetic fields is confined in the
jet (directed flow) and the temperature {\it cannot} be high enough to
produce $^{56}$Ni even for the most energetic model in their study
($B=3\times10^{15}$ G and $\Omega\approx 6000$ rad s$^{-1}$).
More recently, MHD simulations with detailed microphysics, which run
from onset of iron-core collapse to the explosion driven by magnetic
fields, showed that the resultant $^{56}$Ni amount was $\lesssim 0.04
M_\odot$ \citep{nish15} for model with $B\sim10^{15}$ G and
$\Omega\approx 3000$ rad s$^{-1}$ \citep[found in][for hydrodynamic
  explanations of their models]{taki09}.
Therefore, $^{56}$Ni amount cannot be amplified even when we take into
account such MHD driven outflow.

There have been some studies that tried to explain the plateau phase
of the early afterglow by the magnetar scenario because the long
lasting activity can be explained by long-living magnetars. This
discriminates magnetar scenario from the collapsar scenario, whose
lifetime is determined by the accretion timescale of the
hyperaccretion flow.  The typical values for $B_p$ and $\Omega_i$ for
long GRBs are $\gtrsim 3\times 10^{14}$ G and $\gtrsim 6\times 10^3$
rad s$^{-1}$ \citep{troj07} and 3.2 -- 12$\times 10^{14}$ G and
1.7-6.3$\times 10^3$ rad s$^{-1}$ \citep{dall11}. These values are far
less than those given by Eq. (\ref{eq:constraint}). Therefore, if
these GRBs are actually driven by a magnetar, we cannot expect the
bright emission of HNe generated by the decay of $^{56}$Ni. When we
observe a GRB accompanying a HN, whose early afterglow can be
explained by a magnetar with not fulfilling the constraint given by
Eq. (\ref{eq:constraint}), we need an additional energy source to
synthesize $^{56}$Ni other than the dipole radiation from magnetars.

Since the magnetar scenario was recently suggested for the central
engine of superluminous supernovae (SLSNe)
\citep[e.g.][]{kase10,woos10,galy12} as well as GRBs, our discussion
is applicable to this class of explosion. For instance, \citet{kase10}
proposed that $B_p\sim 5\times 10^{14}$ G and $\Omega_i\sim 10^2$ --
$10^3$ rad s$^{-1}$ are required to power the light curve of SLSNe.
Thus, if the magnetar powers SLSNe, the synthesis of $^{56}$Ni, i.e.,
$^{56}$Fe, is not expected. This is contrast to a pair-instability
supernova that is an alternative model for SLSNe.

\section*{Acknowledgements}

YS thanks E. M\"uller for comments and M. Suwa for proofreading. This
study was supported in part by the Grant-in-Aid for Scientific
Research (Nos. 25103511 and 23740157). YS was supported by JSPS
postdoctoral fellowships for research abroad, MEXT SPIRE, and
JICFuS. TN was supported by World Premier International Research
Center Initiative (WPI Initiative), MEXT, Japan.

\appendix
\section{Test calculations}
\label{sec:appendix}

\begin{figure}
\centering
\includegraphics[width=0.4\textwidth]{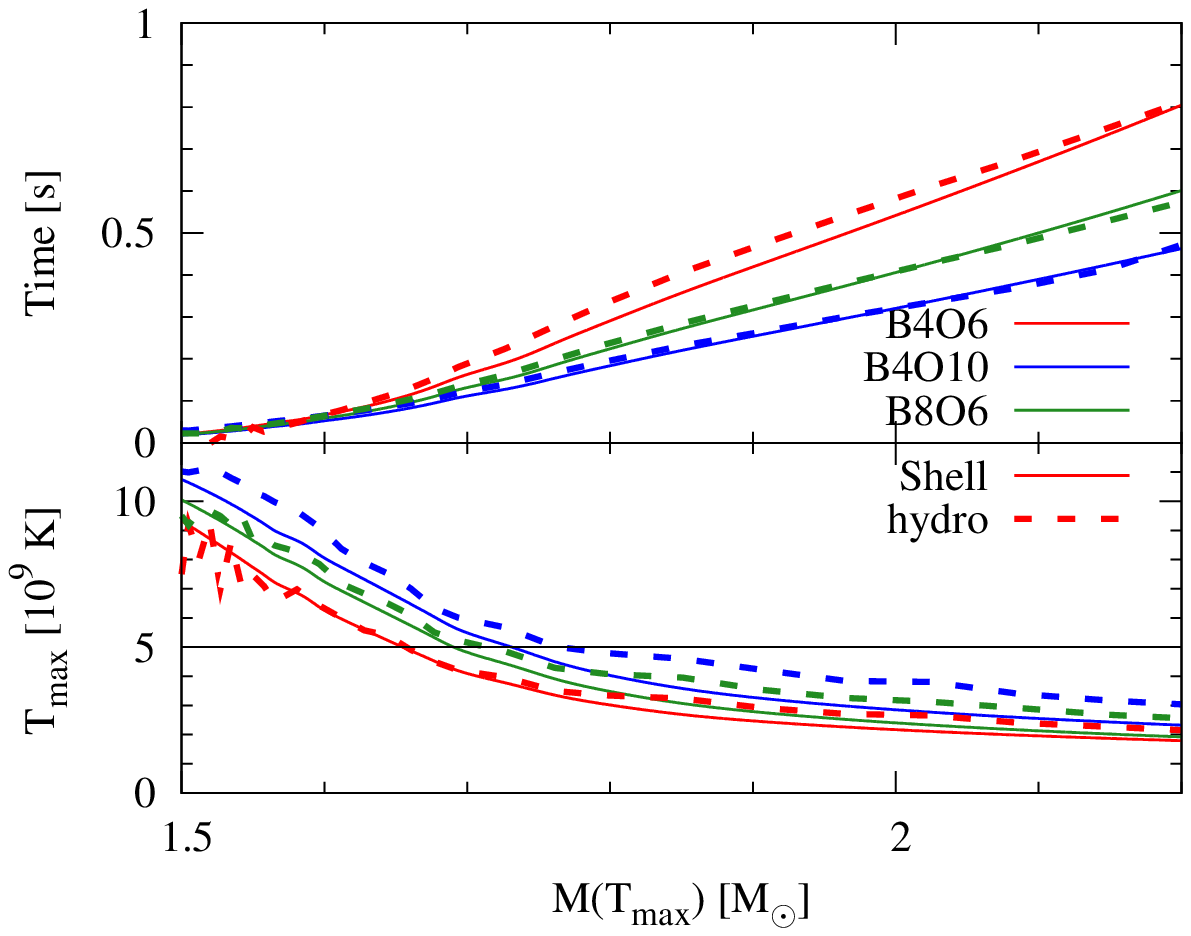}
\caption{The passing time (top panel) and maximum temperature (bottom
  panel) for the shock as a function of mass coordinate. Solid and
  dotted curves represent the results of shell approximation (this
  work) and a hydrodynamic simulation, respectively. Colors represent
  magnetar parameters, $B=4\times10^{16}$ G and $\Omega_i=6000$ rad
  s$^{-1}$ (red), $B=4\times10^{16}$ G and $\Omega_i=10^4$ rad
  s$^{-1}$ (blue), and $B=8\times10^{16}$ G and $\Omega_i=6000$ rad
  s$^{-1}$ (green).  The horizontal dashed line in the bottom panel
  represents the critical temperature for $^{56}$Ni synthesis,
  $5\times 10^9$ K.}
\label{fig:M-T_test}
\end{figure}

Here, we show the validity of our calculation by comparing our
calculation with a hydrodynamic simulation.  In this comparison, we
employ magnetars with three different sets of $B$ and $\Omega_i$,
injected at $M(r)=1.45 M_\odot$ of the 20$M_\odot$ progenitor of
\citet{umed05}.  In Figure \ref{fig:M-T_test}, we show the comparison
of the passing time (top panel) and the maximum temperature (bottom
panel) as a function of mass coordinate for the shell calculation and
the hydrodynamic simulation \citep{tomi07}. The shock and temperature
evolutions computed with these different methods agree quite well and
the systematic error of our thin shell approximation for $^{56}$Ni
mass is $\sim O(0.01) M_\odot$, which is smaller than the
characteristic amount of $^{56}$Ni of HNe, $O(0.1)M_\odot$.

\newcommand\aap{{A\&A}}%
\newcommand\apjl{{ApJ}}%
\newcommand\apj{{ApJ}}%
\newcommand\apjs{{ApJS}}%
\newcommand\mnras{{MNRAS}}%
\newcommand\prc{{Phys. Rev. C}}%
\newcommand\prd{{Phys. Rev. D}}%
\newcommand\pasj{{PASJ}}%
\newcommand\nat{{Nature}}%
\newcommand\araa{{ARA\&A}}%
\newcommand\physrep{{Phys.~Rep.}}%


\begin{thebibliography}{42}
\expandafter\ifx\csname natexlab\endcsname\relax\def\natexlab#1{#1}\fi

\bibitem[{Blinnikov} et~al.(2000){Blinnikov}, {Lundqvist}, {Bartunov}, {Nomoto}
  \& {Iwamoto}]{blin00}
{Blinnikov} S., {Lundqvist} P., {Bartunov} O., {Nomoto} K., {Iwamoto} K., 2000,
  \apj, 532, 1132

\bibitem[{Bucciantini} et~al.(2009){Bucciantini}, {Quataert}, {Metzger},
  {Thompson}, {Arons} \& {Del Zanna}]{bucc09}
{Bucciantini} N., {Quataert} E., {Metzger} B.~D., {Thompson} T.~A., {Arons} J.,
  {Del Zanna} L., 2009, \mnras, 396, 2038

\bibitem[{Burrows} et~al.(2007){Burrows}, {Dessart}, {Livne}, {Ott} \&
  {Murphy}]{burr07b}
{Burrows} A., {Dessart} L., {Livne} E., {Ott} C.~D., {Murphy} J., 2007, \apj,
  664, 416

\bibitem[{Dall'Osso} et~al.(2011){Dall'Osso}, {Stratta}, {Guetta}, {Covino},
  {De Cesare} \& {Stella}]{dall11}
{Dall'Osso} S., {Stratta} G., {Guetta} D., {Covino} S., {De Cesare} G.,
  {Stella} L., 2011, \aap, 526, A121

\bibitem[{Dessart} et~al.(2008){Dessart}, {Burrows}, {Livne} \& {Ott}]{dess08}
{Dessart} L., {Burrows} A., {Livne} E., {Ott} C.~D., 2008, \apjl, 673, L43

\bibitem[{Freiburghaus} et~al.(1999){Freiburghaus}, {Rembges}, {Rauscher}
  et~al.]{frei99}
{Freiburghaus} C., {Rembges} J.-F., {Rauscher} T., et~al., 1999, \apj, 516, 381

\bibitem[{Gal-Yam}(2012)]{galy12}
{Gal-Yam} A., 2012, Science, 337, 927

\bibitem[{Hjorth} \& {Bloom}(2012)]{hjor12}
{Hjorth} J., {Bloom} J.~S., 2012, {The Gamma-Ray Burst - Supernova Connection},
   169--190

\bibitem[{Kasen} \& {Bildsten}(2010)]{kase10}
{Kasen} D., {Bildsten} L., 2010, \apj, 717, 245

\bibitem[{Komissarov} \& {Barkov}(2007)]{komi07}
{Komissarov} S.~S., {Barkov} M.~V., 2007, \mnras, 382, 1029

\bibitem[{Koo} \& {McKee}(1990)]{koo90}
{Koo} B.-C., {McKee} C.~F., 1990, \apj, 354, 513

\bibitem[{Laumbach} \& {Probstein}(1969)]{laum69}
{Laumbach} D.~D., {Probstein} R.~F., 1969, Journal of Fluid Mechanics, 35, 53

\bibitem[{LeBlanc} \& {Wilson}(1970)]{lebl70}
{LeBlanc} J.~M., {Wilson} J.~R., 1970, \apj, 161, 541

\bibitem[{MacFadyen} \& {Woosley}(1999)]{macf99}
{MacFadyen} A.~I., {Woosley} S.~E., 1999, \apj, 524, 262

\bibitem[{Maeda} et~al.(2002){Maeda}, {Nakamura}, {Nomoto}, {Mazzali}, {Patat}
  \& {Hachisu}]{maed02}
{Maeda} K., {Nakamura} T., {Nomoto} K., {Mazzali} P.~A., {Patat} F., {Hachisu}
  I., 2002, \apj, 565, 405

\bibitem[{Maeda} \& {Tominaga}(2009)]{maed09}
{Maeda} K., {Tominaga} N., 2009, \mnras, 394, 1317

\bibitem[{Meier} et~al.(1976){Meier}, {Epstein}, {Arnett} \& {Schramm}]{meie76}
{Meier} D.~L., {Epstein} R.~I., {Arnett} W.~D., {Schramm} D.~N., 1976, \apj,
  204, 869

\bibitem[{Metzger} et~al.(2011){Metzger}, {Giannios}, {Thompson}, {Bucciantini}
  \& {Quataert}]{metz11}
{Metzger} B.~D., {Giannios} D., {Thompson} T.~A., {Bucciantini} N., {Quataert}
  E., 2011, \mnras, 413, 2031

\bibitem[{M{\"o}sta} et~al.(2014){M{\"o}sta}, {Richers}, {Ott} et~al.]{moes14}
{M{\"o}sta} P., {Richers} S., {Ott} C.~D., et~al., 2014, \apjl, 785, L29

\bibitem[{Nagataki} et~al.(2006){Nagataki}, {Mizuta} \& {Sato}]{naga06}
{Nagataki} S., {Mizuta} A., {Sato} K., 2006, \apj, 647, 1255

\bibitem[{Nakamura} et~al.(2001{\natexlab{a}}){Nakamura}, {Mazzali}, {Nomoto}
  \& {Iwamoto}]{naka01c}
{Nakamura} T., {Mazzali} P.~A., {Nomoto} K., {Iwamoto} K., 2001{\natexlab{a}},
  \apj, 550, 991

\bibitem[{Nakamura} et~al.(2001{\natexlab{b}}){Nakamura}, {Umeda}, {Iwamoto}
  et~al.]{naka01b}
{Nakamura} T., {Umeda} H., {Iwamoto} K., et~al., 2001{\natexlab{b}}, \apj, 555,
  880

\bibitem[Nishimura et al.(2015)]{nish15} Nishimura, N., 
Takiwaki, T., \& Thielemann, F.-K.\ 2015, arXiv:1501.06567 

\bibitem[{Nomoto} et~al.(2006){Nomoto}, {Tominaga}, {Tanaka} et~al.]{nomo06}
{Nomoto} K., {Tominaga} N., {Tanaka} M., et~al., 2006, Nuovo Cimento B Serie,
  121, 1207

\bibitem[{Sawai} et~al.(2013){Sawai}, {Yamada} \& {Suzuki}]{sawa13}
{Sawai} H., {Yamada} S., {Suzuki} H., 2013, \apjl, 770, L19

\bibitem[{Shapiro} \& {Teukolsky}(1983)]{shap83}
{Shapiro} S.~L., {Teukolsky} S.~A., 1983, {Black holes, white dwarfs, and
  neutron stars: The physics of compact objects}, New York, Wiley-Interscience,
  1983, 663 p.

\bibitem[{Symbalisty}(1984)]{symb84}
{Symbalisty} E.~M.~D., 1984, \apj, 285, 729

\bibitem[{Takiwaki} et~al.(2009){Takiwaki}, {Kotake} \& {Sato}]{taki09}
{Takiwaki} T., {Kotake} K., {Sato} K., 2009, \apj, 691, 1360

\bibitem[{Thompson} et~al.(2004){Thompson}, {Chang} \& {Quataert}]{thom04}
{Thompson} T.~A., {Chang} P., {Quataert} E., 2004, \apj, 611, 380

\bibitem[{Tominaga}(2009)]{tomi09}
{Tominaga} N., 2009, \apj, 690, 526

\bibitem[{Tominaga} et~al.(2007){Tominaga}, {Maeda}, {Umeda} et~al.]{tomi07}
{Tominaga} N., {Maeda} K., {Umeda} H., et~al., 2007, \apjl, 657, L77

\bibitem[{Troja} et~al.(2007){Troja}, {Cusumano}, {O'Brien} et~al.]{troj07}
{Troja} E., {Cusumano} G., {O'Brien} P.~T., et~al., 2007, \apj, 665, 599

\bibitem[{Umeda} \& {Nomoto}(2005)]{umed05}
{Umeda} H., {Nomoto} K., 2005, \apj, 619, 427

\bibitem[{Usov}(1992)]{usov92}
{Usov} V.~V., 1992, \nat, 357, 472

\bibitem[{Whitworth} \& {Francis}(2002)]{whit02}
{Whitworth} A.~P., {Francis} N., 2002, \mnras, 329, 641

\bibitem[{Winteler} et~al.(2012){Winteler}, {K{\"a}ppeli}, {Perego}
  et~al.]{wint12}
{Winteler} C., {K{\"a}ppeli} R., {Perego} A., et~al., 2012, \apjl, 750, L22

\bibitem[{Woosley}(1993)]{woos93}
{Woosley} S.~E., 1993, \apj, 405, 273

\bibitem[{Woosley}(2010)]{woos10}
{Woosley} S.~E., 2010, \apjl, 719, L204

\bibitem[{Woosley} \& {Bloom}(2006)]{woos06}
{Woosley} S.~E., {Bloom} J.~S., 2006, \araa, 44, 507

\bibitem[{Woosley} \& {Heger}(2007)]{woos07}
{Woosley} S.~E., {Heger} A., 2007, \physrep, 442, 269

\bibitem[{Woosley} et~al.(2002){Woosley}, {Heger} \& {Weaver}]{woos02}
{Woosley} S.~E., {Heger} A., {Weaver} T.~A., 2002, Reviews of Modern Physics,
  74, 1015

\bibitem[{Yoshida} et~al.(2014){Yoshida}, {Okita} \& {Umeda}]{yosh14}
{Yoshida} T., {Okita} S., {Umeda} H., 2014, \mnras, 438, 3119

\end{thebibliography}
\end{document}